\documentclass[prd,reprint,showpacs,nofootinbib]{revtex4-1}
\usepackage{array,graphicx,multirow,amsmath,amsfonts,mathrsfs}
\usepackage{nicefrac}
\usepackage[T1]{fontenc}
\usepackage{tikz}
\usetikzlibrary{tikzmark}

\newcommand{\re}[1]{(\ref{eq:#1})}

\def\phi{\varphi}

\def\rho{\varrho}
\def\d{\mathrm{d}}
\def\p{\partial}
\def\dAl{\Box}

\renewcommand{\vec}[1]{\boldsymbol{#1}}

\newcommand{\Lie}[1]{\vec{\mathcal{L}}_{\vec{#1}}}

\newcommand{\rem}[1]{}

\def\={\discretionary{-}{-}{-}}

{\vskip 3pt plus 1pt minus 1pt\begin{normalsize}}%
{\par\end{normalsize}\vskip 3pt plus 1pt minus 1pt}

\newlength{\obrA} 
\newlength{\obrB} 

\newcommand{\Scri}{\mathscr{S}}
\newcommand{\Epsilon}{\mathcal{E}}

\newcommand{\mth}{\text{\th}}
\newcommand{\mdh}{\text{\dh}}

\newcommand{\Tphi}{\tilde{T}}
\newcommand{\Xphi}{\tilde{X}}
\newcommand{\Yphi}{\tilde{Y}}
\newcommand{\Xtau}{\hat{X}}
\newcommand{\Ytau}{\hat{Y}}
\newcommand{\Ptau}{\hat{P}}

\newcommand{\Ktau}{K_\tau}
\newcommand{\Kphi}{K_{\!\phi}}

\newcommand{\Rp}{r_{\!p\,}}
\newcommand{\Rm}{r_m}

\newcommand{\pFq}[4]{{}_{2}F_{1}{\left(\genfrac..{0pt}{}{#1,\,#2}{#3};\, #4\right)}}

\begin{document}

\title{Separability of test fields equations on the C\,--\,metric background}

\date{\today}
\author{David Kofro\v{n}}
\email{d.kofron@gmail.com}
\affiliation{
Institute of Theoretical Physics, Faculty of Mathematics and Physics,\\
Charles University in Prague,\\
V Hole\v{s}ovi\v{c}k\'{a}ch 2, 180\,00 Prague 8, Czech Republic}


\begin{abstract}
In the Kerr\,--\,Newman spacetime the Teukolsky master equation, governing the fundamental test fields, is of great importance. We derive an analogous master equation for the non-rotating C\,--\,metric which encompass massless Klein\,--\,Gordon field, neutrino field, Maxwell field, Rarita\,--\,Schwinger field and gravitational perturbations. This equation is shown to be separable in terms of ``accelerated spin weighted spherical harmonics''. It is shown that, contrary to ordinary spin weighted spherical harmonics, the ``accelerated'' ones are different for different spins. In some cases, the equation for eigenfunctions and eigenvalues are explicitly solved.
\end{abstract}
\maketitle

\section{Introduction}

Boost\,--\,rotation symmetric spacetimes are a special class of solutions of vacuum Einstein's field equations with two symmetries which can represent moving objects and contain radiation. Some exact explicit solutions belonging to this class are e.g. the C\,--\,metric \cite{weyl} and the solutions found by Bonnor and Swaminarayan \cite{bszp}. In general these spacetimes are algebraically general, radiative as shown in \cite{bi68} and possess a plausible Newtonian limit \cite{BiKofNL}.

Among these solutions the C\,--\,metric is a special case. It is of Petrov type D --- and thus  a generalization \cite{pd} admitting charge and rotation could have been found, involving ``uniformly accelerated black holes''.

The method of decoupling and separating variables of the equations governing fundamental fields is, of course, of great importance but it has not been applied to the C\,--\,metric so far. 

In general relativity, the decoupling and separation of variables is often employing the Newman\,--\,Penrose (NP) formalism. If NP formalism is applied to the type D spacetimes, the equations for radiative (ingoing and outgoing radiation) components of (a) massless Klein-Gordon field ($s=0$), (b) neutrino field ($s=\nicefrac{1}{2}$), (c) test Maxwell field ($s=1$), (d) Rarita\,--\,Schwinger field ($s=\nicefrac{3}{2}$) and (e) linear gravitational perturbations ($s=2$) can be decoupled \cite{Teuk}, \cite{RS}, and in the case of Kerr\,--\,Newman metric they can be also separated. We rewrite these equations in Geroch\,--\,Held\,--\,Penrose (GHP) formalism \cite{sc} in Section \ref{sec:meq}. In Section \ref{sec:proj}, we review some elementary concepts of the Ernst projection formalism which we later connect to the NP formalism. 

The background metric -- the C\,--\,metric -- is presented in Section \ref{sec:cm}. For more details see \cite{kw}, \cite{ht1}, \cite{gkp} which encompass comprehensive historical introduction. The equation analogous to the Teukolsky master equation is obtained in Section \ref{sec:meq} for the charged C\,--\,metric. We show that the solution of this equation can be found using separation of variables in the canonical coordinates \cite{ht1}.  The angular part of the solution leads to generalized ``accelerated spin weighted spherical harmonics''\footnote{This terminology can be disputed since these solutions are not in general eigenfunctions of Laplace's operator.}. These are in general Heun \cite{heun} functions, but for extremal case the solutions reduce to rational functions. We find the eigenfunctions and eigenvalues for axially symmetric fields ($m=0$) in Section \ref{sec:ang} and show why for $m$ different from zero this is difficult, solving the same problem for cosmic string spacetime (the C\,--\,metric inherently contains deficit angle) in Section \ref{sec:cs}.  

The electromagnetic field which was the primary motivation is then analysed in a more detailed way in section \ref{sec:elmag}.

Contrary to ordinary spin weighted spherical harmonics the accelerated ones form more complicated structure as they are split into different families according  to the spin $S$ of the field, this issue is discussed in section \ref{sec:sw}.

\section{Ernst projection formalism}\label{sec:proj}

A well known fact is that in vacuum spacetimes every Killing vector can serve as a 4-potential of a test electromagnetic field. The straightforward generalization to spacetimes with matter can be done with the help of the Ernst projection formalism.

In the presence of the Killing vector field $\vec{\xi}$, the Ernst projection formalism \cite{ernst-axi}, \cite{ernst-axii} can be used to solve the full Einstein\,--\,Maxwell equations.
But the Ernst equations can be used for solving Maxwell equations for test electromagnetic field (which respects the symmetry induced by the aforementioned Killing vector field, i.e. $\Lie{\xi} \vec{F} = 0$) on a given background.

Let $\vec{F}$ be an electromagnetic field for which $\Lie{\xi} \vec{F} = 0$, then the potential $\Psi$ with respect to the Killing vector $\vec{\xi}$ can be introduced\footnote{Complex self-dual 2-form is defined as $\vec{F}^*=\vec{F}+\frac{i}{2}\,\vec{\star F}$.}
\begin{align}
\sqrt{\frac{\kappa}{2}}\;\xi^a\,F^*_{ab} &=\Psi_{,b}\,, &
\Psi_{,a} \xi^a &=0 \,,&
F^{*ab}_{\phantom{*ab};b} &=0\,,
\end{align}
and, conversely,
\begin{equation}
\sqrt{\frac{\kappa}{2}}\,F^*_{ab} = 2 F^{-1}\left( \xi_{[a}\Psi_{,b]} \right)^*\,.
\label{eq:Fpot}
\end{equation}
The inverse of the norm of the Killing vector in this equation is the reason that only very special solutions with respect to general linear combination of timelike and spacelike KV are, though mathematically possible, physically relevant.
Covariant divergence of \re{Fpot} gives the equation for $\Psi$ 
\begin{equation}
F\left( \frac{\Psi_{,a}}{F} \right)^{;a} + i\omega^a\left( \frac{\Psi_{,a}}{F} \right) = 0\,.
\label{eq:PsiEq}
\end{equation}
which is a \emph{linear} PDE where
\begin{align}
F &= \xi^a\xi_a, & \omega^a &= \varepsilon^{abcd}\,\xi_b\,\xi_{c;d}\,.
\end{align}

The equation \re{PsiEq} is a mid-step in derivation of Ernst equations (for details presented in modern fashion see \cite{sc}) where the norm $F$ of KV $\vec{\xi}$ is identified with $-F = \frac{1}{2}\left( \Epsilon+\bar{\Epsilon} \right)+\bar{\Phi}{\Phi}$, where $\Epsilon$ is so called the gravitational Ernst potential whereas $\Phi$ is the electromagnetic Ernst potential
\begin{eqnarray}
F\, \dAl \Epsilon &=& \left( \nabla_a \Epsilon + 2\bar{\Phi}\nabla_a \Phi \right)\nabla^a \Epsilon\,,\\
F\, \dAl \Phi &=& \left( \nabla_a \Epsilon + 2\bar{\Phi}\nabla_a \Phi \right)\nabla^a \Phi\,.
\end{eqnarray}
For a given spacetime the Ernst potentials can be calculated. And these two potentials are solutions of Maxwell equations for test field \re{PsiEq} with $\Psi=\Phi$ or $\Psi=\Epsilon$.

\section{Non-rotating C\,--\,metric}\label{sec:cm}
The non-rotating C\,--\,metric, a space-time representing two\footnote{After the analytic continuation across the acceleration horizon.} in general charged uniformly accelerated black holes \cite{kw}, \cite{gkp}, is given in a modern form \cite{ht1} with factorized structure function $G(\xi)$ by
\begin{multline}
\vec{\d} s^2 = \frac{1}{A^2\left( x-y \right)^2}\Biggl[ \Ktau^2 G(y)\,\vec{\d}\tau^2 -\frac{1}{G(y)}\,\vec{\d} y^2 + \\
+ \frac{1}{G(x)}\,\vec{\d} x^2 + \Kphi^2\, G(x)\,\vec{\d}\phi^2 \Biggr]\,. \label{eq:cm} 
\end{multline}
The structure function $G(\xi)$ is given by the fourth order polynomial
\begin{equation}
G(\xi) = \left( 1-\xi^2 \right)\left( 1+A\Rp \xi \right)\left( 1+A\Rm\xi \right)\,, \label{eq:G}
\end{equation}
in which the structure of roots is defined by $-\infty < -1/A\Rm < -1/A\Rp < -1$. The parameter $A$ defines the acceleration of the black hole and parameters $\Rp$, $\Rm$ can be related to mass parameter $M$ and charge\footnote{For the sake of completeness, let us mention that the four potential of electromagnetic field is given by $\vec{A}= \Ktau q y\,\vec{d}\tau$.} parameter $q$ by
\begin{align}
\Rp &= M+\sqrt{M^2-q^2}\,, & \Rm &=M-\sqrt{M^2-q^2}\,.
\end{align}

The metric \re{cm} covers several regions of the spacetime but we are interested in the asymptotically flat region outside the black hole which is covered by coordinate ranges $\tau\in \mathbb{R}$, $y\in \langle-1/A\Rp,\,1\rangle$ (with $y=-1/A\Rp$ being the black hole horizon and $y=-1$ being the acceleration horizon), $x\in\langle 1,\,-1\rangle$ (axis given by $x=\pm 1$) and $x-y<0$ (the asymptotic infinity is located at $x-y=0$) and, finally,  $\phi \in \langle 0,\,2\pi\rangle$. The strength of the conical singularity which is inevitably present in the C\,--\,metric is governed by the parameter $\Kphi$. The constant $\Ktau$ can, of course, be absorbed in definition of $\tau$ but we leave it explicitly present. 

The null tetrad adapted to the principal null directions\footnote{It is chosen such that in the limit $\Rp=\Rm=0$, i.e. in the flat spacetime where $G(\xi)=1-\xi^2$, and after standard coordinate transformation $x=\cos\theta$ the vector $\vec{m}$ becomes $\vec{\p_\theta}+\frac{i}{\sin\theta}\,\vec{\p_\phi}$.} reads
\begin{align}
\vec{l} &= -\frac{A^2\left( x-y \right)^2}{\sqrt{2}}\, \left( \frac{1}{G(y)\Ktau}\,\vec{\frac{\p}{\p \tau}} - \vec{\frac{\p}{\p y}} \right), \\
\vec{n} &= -\frac{G(y)}{\sqrt{2}}\left( \frac{1}{G(y)\Ktau}\,\vec{\frac{\p}{\p \tau}}  -\vec{\frac{\p}{\p y}}\right), \\ 
\vec{m} &= \frac{1}{\sqrt{2}}\frac{A(x-y)}{\sqrt{G(x)}} \left(-G(x)\,\vec{\frac{\p}{\p x}} + \frac{i}{\Kphi}\,\vec{ \frac{\p}{\p\phi}}\right),\\
\vec{\bar{m}} &= \frac{1}{\sqrt{2}}\frac{A(x-y)}{\sqrt{G(x)}} \left(-G(x)\,\vec{\frac{\p}{\p x}} - \frac{i}{\Kphi}\,\vec{ \frac{\p}{\p\phi}}\right),
\label{eq:PND}
\end{align}
and the corresponding nonzero NP coefficients are
\begin{align}
\pi & = \frac{1}{\sqrt{2}}\,A\sqrt{G(x)}\,, \\
\mu & = \frac{1}{\sqrt{2}}\frac{G(y)}{x-y}\,, \\
\tau &= -\frac{1}{\sqrt{2}}\,A\sqrt{G(x)}\,, \\
\rho &= -\frac{1}{\sqrt{2}}\, A^2\left( x-y \right) \,, \\
\gamma &= -\frac{\sqrt{2}}{4}\left[ \frac{G(y)}{(x-y)^2} \right]_{,y}\left( x-y \right)^2\,, \\
\beta &= -\frac{\sqrt{2}}{4}\,A(x-y)\left( \sqrt{G(x)} \right)_{,x}\,, \\
\alpha &= \frac{A\sqrt{2}}{4}\,\left[ \frac{\sqrt{G(x)}}{(x-y)^2} \right]_{,x}\left( x-y \right)^3\,. 
\label{eq:sc}
\end{align}
Finally, the only non-zero Weyl NP scalar is
\begin{equation}
\psi_2 = - A^3\left( x-y \right)^3 \left( \frac{\Rp+\Rm}{2} +A\Rp\Rm\left( x+y \right) \right)\,.
\end{equation}

Let us assume that we have a potential $\Psi^{(\tau)}(y,\,x,\,\phi)$ of \re{PsiEq} with respect to the boost Killing vector $\vec{\xi_{(\tau)}}$, resp. $\Psi^{(\phi)}(\tau,\,y,\,x)$ with respect to the axial Killing vector $\vec{\xi_{(\phi)}}$. The equation \re{PsiEq} for Ernst electromagnetic potential on the C\,--\,metric background is separable in both cases, i.e. let us assume that $\Psi^{(\tau)} = \Ytau(y)\Xtau(x)\Ptau(\phi)$ and $\Psi^{(\phi)}= \Tphi(\tau)\Yphi(y)\Xphi(x)$. Then separated equations read
\begin{eqnarray}
G(y)\Ytau_{,yy} +  \Lambda \Ytau & = &0\,,\\
\left[ G(x)\Xtau_{,x} \right]_{,x} + \left[ \Lambda - \frac{m^2}{\Kphi^2 G(x)}\right] \Xtau &=&0\,, \\
\Ptau_{,\phi\phi} + m^2 \Ptau & = &0\,; \label{eq:EPsep1}
\end{eqnarray}
and
\begin{eqnarray}
\Tphi_{,\tau\tau} + \omega^2\Tphi &=& 0\,, \\
\left[ G(y)\Yphi_{,y} \right]_{,y} + \left[ \Lambda + \frac{\omega^2}{\Ktau^2 G(y)} \right]\Yphi &=&0\,, \\
G(x)\Xphi_{,xx} + \Lambda \Xphi &=&0\,.
\label{eq:EPsep2}
\end{eqnarray}

From the Ernst potential $\Psi^{(\cdot)}$ we can construct the electromagnetic field tensor $\vec{F}$ by \re{Fpot} and calculate the NP scalars of electromagnetic field corresponding to potential $\Psi^{(\phi)}$, resp. $\Psi^{(\tau)}$. They are given by
\begin{align}
\Phi^{(\phi)}_0 &= \frac{-i\sqrt{2}A^3\left( x-y \right)^3}{\Kphi\Ktau}\,\frac{\left( \Ktau G(y)\Psi^{(\phi)}_{,y} - \Psi^{(\phi)}_{,\tau} \right)}{G(y)\sqrt{G(x)}}\,,\\
\Phi^{(\phi)}_1 &= \frac{i\sqrt{2}\,A^2(x-y)^2 \Psi^{(\phi)}_{,x}}{\Kphi}\,, \\
\Phi^{(\phi)}_2 &= \frac{i\sqrt{2} A( x-y)}{\Kphi\Ktau}\,\frac{\left( \Ktau G(y)\Psi^{(\phi)}_{,y} + \Psi^{(\phi)}_{,\tau} \right)}{\sqrt{G(x)}}\,,
\label{eq:EPs1}
\end{align}
and
\begin{align}
\Phi^{(\tau)}_0 &= \frac{-\sqrt{2}iA^3\left( x-y \right)^3}{\Kphi\Ktau}\,\frac{\left( \Kphi G(x)\Psi^{(\tau)}_{,x} - i\Psi^{(\tau)}_{,\phi} \right)}{G(y)\sqrt{G(x)}}\,, \label{eq:tau0}\\
\Phi^{(\tau)}_1 &= \frac{\sqrt{2}\,A^2(x-y)^2 \Psi^{(\phi)}_{,y}}{\Kphi}\,, \label{eq:tau1}\\
\Phi^{(\tau)}_2 &= \frac{i\sqrt{2}A(x-y)}{\Kphi\Ktau}\,\frac{\left( \Kphi G(x)\Psi^{(\tau)}_{,x} + i\Psi^{(\tau)}_{,\phi} \right)}{\sqrt{G(x)}}\,, \label{eq:tau2}
\end{align}

We observe that there exists a class of solutions for which $\Lie{\xi_{(\phi)}} \vec{F} = \Lie{\xi_{(\tau)}} \vec{F} = 0$ at the same time and thus we can find both potentials for this field. This gives us a relation between solutions of equations \re{EPsep1} and \re{EPsep2}. If $\mathcal{V}$ is solution of $\left(G\mathcal{V}_{,z}\right)_{,z}+\lambda \mathcal{V}=0$ then $\mathcal{U}=G\mathcal{V}_{,z}$ is solution of the equation $G\mathcal{U}_{,zz}+\lambda\mathcal{U}=0$. And, conversely, let $\mathcal{U}$ be the solution of $G\mathcal{U}_{,zz}+\lambda\mathcal{U}=0$ then $\mathcal{V}=\mathcal{U}_{,z}$ is solution of $\left(G\mathcal{V}_{,z}\right)_{,z}+\lambda \mathcal{V}=0$.

More generally, the Ernst potential can be found for a general linear combination of $\vec{\xi_{(\tau)}}$ and $\vec{\xi_{(\phi)}}$ if the Lie derivative of the electromagnetic field along this vector field vanishes, but then the separation of variables leads to static and axially symmetric solutions only.

\section{Master equation}\label{sec:meq}
Teukolsky \cite{Teuk} provided decoupled equations for NP components of gravitational perturbation $\Psi_0$ and $\Psi_4$, of test electromagnetic field $\Phi_0$ and $\Phi_2$ and neutrino field $\chi_0$ and $\chi_1$ in general type D spacetime and performed a detailed analysis of these equations on the Kerr\,--\,Newman background. The C\,--\,metric is also a type D solution and therefore we can perform a similar analysis.  The equations\footnote{We consider vacuum solutions only.} for gravitational perturbations reads
\begin{eqnarray}
\bigl[ \left( D - 3\epsilon+\bar{\epsilon} -4\varrho -\bar{\varrho} \right)\left( \Delta-4\gamma+\mu \right)\qquad\qquad\qquad && \label{eq:Psi0}\\ - \left( \delta +\bar{\pi}-\bar{\alpha}-3\beta-4\tau \right)\left( \bar{\delta}+\pi-4\alpha \right)-3\psi_2 \bigr] \Psi_0 &=& 0\,,
\nonumber \\
\bigl[ \left( \Delta +3\gamma-\bar{\gamma}+4\mu+\bar{\mu} \right)\left( D+4\epsilon-\varrho \right)\qquad\qquad\qquad && \label{eq:Psi4}\\- \left( \bar{\delta} -\bar{\tau}+\bar{\beta}+3\alpha+4\pi \right)\left( \delta-\tau+4\beta \right) -3\psi_2 \bigr] \Psi_4 &=& 0\,, \nonumber
\end{eqnarray}
Rarita\,--\,Schwinger equation \cite{RS}
\begin{eqnarray}
\bigl[ \left( D - 2\epsilon+\bar{\epsilon} -3\varrho -\bar{\varrho} \right)\left( \Delta-3\gamma+\mu \right) \qquad\qquad\qquad&& \label{eq:RS0} \\- \left( \delta +\bar{\pi}-\bar{\alpha}-2\beta-3\tau \right)\left( \bar{\delta}+\pi-3\alpha \right)-\psi_2 \bigr] \Sigma_0 &=& 0\,, \nonumber \\
\bigl[ \left( \Delta +2\gamma-\bar{\gamma}+3\mu+\bar{\mu} \right)\left( D+3\epsilon-\varrho \right) \qquad\qquad\qquad &&\label{eq:RS3} \\ - \left( \bar{\delta} -\bar{\tau}+\bar{\beta}+2\alpha+3\pi \right)\left( \delta-\tau+3\beta \right) -\psi_2 \bigr] \Sigma_3 &=& 0\,, \nonumber
\end{eqnarray}
for test electromagnetic field
\begin{eqnarray}
\bigl[ \left( D-\epsilon+\bar{\epsilon}-2\varrho-\bar{\varrho} \right)\left( \Delta-2\gamma+\mu \right)\qquad\qquad\qquad \label{eq:Phi0} \\- \left( \delta +\bar{\pi} -\bar{\alpha}-\beta-2\tau\right)\left( \bar{\delta}+\pi-2\alpha \right) \bigr]\Phi_0 &=&0\,, \nonumber\\
\bigl[ \left( \Delta+\gamma-\bar{\gamma}+2\mu+\bar{\mu} \right)\left( D+2\epsilon-\varrho \right)\qquad\qquad\qquad && \label{eq:Phi2}\\ - \left( \bar{\delta}-\bar{\tau}+\bar{\beta}+\alpha+2\pi \right)\left( \Delta-\tau+2\beta \right) \bigr]\Phi_2 &=&0 \,, \nonumber
\end{eqnarray}
and, neutrino equation
\begin{eqnarray}
\bigl[ \left( D+\bar{\epsilon}-\varrho-\bar{\varrho} \right)\left( \Delta-\gamma+\mu \right)\qquad\qquad\qquad &&\label{eq:chi0} \\- \left( \delta+\bar{\pi}-\bar{\alpha}-\tau \right)\left( \bar{\delta}+\pi-\alpha \right) \bigr]\chi_0 &=& 0\,, \nonumber \\
\bigl[ \left( \Delta-\bar{\gamma}+\mu+\bar{\mu} \right)\left( D+\epsilon-\varrho \right)\qquad\qquad\qquad && \label{eq:chi1}\\- \left( \bar{\delta}-\bar{\tau}+\bar{\beta}+\pi \right)\left( \delta-\tau+\beta \right) \bigr]\chi_1 &=&0 \,,\nonumber
\end{eqnarray}

Equations \re{Psi0}, \re{RS0}, \re{Phi0} and \re{chi0} for NP scalars can be rewritten using the GHP formalism in more compact form: 
\begin{eqnarray}
\bigl[ \left( \mth -\bar{\rho}-4\rho \right)\left( \mth' +\mu \right)\qquad\qquad\qquad &&\\ - \left( \mdh+\bar{\pi}-4\tau \right)\left( \mdh'+\pi \right) -3\psi_2 \bigr] \Psi_0 & =& 0 \qquad \text{for } s=2 \,,\nonumber \\
\bigl[ \left( \mth -\bar{\rho}-3\rho \right)\left( \mth' +\mu \right)\qquad\qquad\qquad&&\\ - \left( \mdh+\bar{\pi}-3\tau \right)\left( \mdh'+\pi \right) -\phantom{3}\psi_2 \bigr] \Sigma_0 & =& 0 \qquad \text{for } s=\nicefrac{3}{2} \,,\nonumber \\
\bigl[ \left( \mth -\bar{\rho}-2\rho \right)\left( \mth' +\mu \right)\qquad\qquad\qquad&& \\ - \left( \mdh+\bar{\pi}-2\tau \right)\left( \mdh'+\pi \right) \phantom{-3\psi_2}\bigr] \Phi_0 & =& 0 \qquad \text{for } s=1 \,, \nonumber \\
\bigl[ \left( \mth -\bar{\rho}-\rho \right)\left( \mth' +\mu \right)\qquad\qquad\qquad &&\\- \left( \mdh+\bar{\pi}-\tau \right)\left( \mdh'+\pi \right) \phantom{-3\psi_2}\bigr] \chi_0 & =& 0 \qquad \text{for } s=\nicefrac{1}{2} \,,\nonumber
\end{eqnarray}
from which we infer, for $\Phi \in (\Psi_0,\,\Sigma_0,\,\Phi_0,\,\chi_0)$ and general $s>0$ 
\begin{multline}
\bigl[ \left( \mth -\bar{\rho}-2s\rho \right)\left( \mth' +\mu \right) - \left( \mdh+\bar{\pi}-2s\tau \right)\left( \mdh'+\pi \right)\\ -(2s-1)(s-1)\psi_2 \bigr] \Phi  = 0 \,, 
\end{multline}
and the same analysis of equations \re{Psi4}, \re{RS3}, \re{Phi2} and \re{chi1} results in equation for $\Phi \in (\Psi_4,\,\Sigma_3,\,\Phi_2,\,\chi_1)$ for general $s<0$
\begin{multline}
\bigl[ \left( \mth'+\bar{\mu}-2s\mu \right)\left( \mth-\rho \right)-\left( \mdh'-\bar{\tau}-2s\pi \right)\left( \mdh-\tau \right)\\ -\left( 2s+1 \right)\left( s+1 \right)\psi_2 \bigr] \Phi = 0\,,
\end{multline}
which leads to massless Klein\,--\,Gordon ($s=0$) equation in GHP formalism
\begin{equation}
\left[ \left( \mth -\bar{\rho} \right)\left( \mth' +\mu \right) - \left( \mdh+\bar{\pi} \right)\left( \mdh'+\bar{\pi} \right)-\psi_2 \right] \Phi  = 0  \,, \label{eq:KG-GHP}
\end{equation}
or, in NP formalism
\begin{multline} 
\bigl[ \left( D +\epsilon+\bar{\epsilon}  -\bar{\varrho} \right)\left( \Delta+\mu \right)\\ - \left( \delta +\bar{\pi}-\bar{\alpha}+\beta \right)\left( \bar{\delta}+\pi \right)-\psi_2 \bigr] \Phi =0\,,
\label{eq:KG-NP}
\end{multline}
which is an assertion we have to prove. 

The d'Alembert operator acting on scalar is simply given by 
\begin{multline}
\Box = \nabla^a \nabla_a = \left( -n_a D - l_a \Delta + \bar{m}_a\delta +m_a\bar{\delta} \right) \times  \\
\left( -n^a D - l^a \Delta + \bar{m}^a\delta +m^a\bar{\delta} \right)\,,
\label{eq:dAl}
\end{multline}
The d'Alembert operator then  can be expressed (cf. \cite{Sch2}) as
\begin{multline}
-\frac{1}{2}\,\Box = D\Delta -\delta\bar{\delta} +\mu D +\left( \epsilon+\bar{\epsilon}-\bar{\rho} \right)\Delta -\phi\delta \\+ \left( \bar{\alpha}-\beta-\bar{\rho} \right)\bar{\delta}\,,
\label{eq:dAl-NPn}
\end{multline}
but expanding $\left[\left( \mth -\bar{\rho} \right)\left( \mth' +\mu \right) - \left( \mdh+\bar{\pi} \right)\left( \mdh'+\pi \right)\right]$ (acting on field of GHP weight (0,0)) and using the Ricci identity for $D\mu - \delta\phi$; or expanding $\left[ \left( \mth'+\bar{\mu} \right)\left( \mth-\rho \right)-\left(\mdh'-\bar{\tau} \right)\left( \mdh-\tau \right) \right]$ and using the Ricci identity for $\Delta\rho - \bar{\delta}\tau$  results in  
\begin{multline}
\left[ \left( \mth -\bar{\rho} \right)\left( \mth' +\mu \right) - \left( \mdh+\bar{\pi} \right)\left( \mdh'+\pi \right) \right] \\ 
=\left[ \left( \mth'+\bar{\mu} \right)\left( \mth-\rho \right)-\left(\mdh'-\bar{\tau} \right)\left( \mdh-\tau \right) \right] \\
= -\frac{1}{2} \left[ \Box -2\psi_2-\frac{4}{3}R -2\sigma\lambda+2\nu\kappa \right]\,. 
\end{multline}
Thus \re{KG-GHP} is indeed the massless Klein\,--\,Gordon equation (for Ricci flat type D spacetimes).

Regarding the NP spin coefficients given in \re{sc}, the equations listed above can be represented as a master equation for $\hat{\Phi}(\tau,\,y,\,x,\,\phi)$ which is a function listed in the table \ref{tab:F} 

\begin{table*}
\centering
\caption{Separable ansatz, spin a GHP weight of field components.}
\renewcommand{\arraystretch}{1.5}
\begin{ruledtabular}
\begin{tabular}{l|ccccccccc} 
$\hat{\Phi}$ & 
$\rho^{-1}\Psi_4$ & $\rho^{-1}\Sigma^{RS}_3$ &
$\rho^{-1}\Phi^{EM}_2$ & $\rho^{-1}\chi_1$ &
$\rho^{-1}\Phi^{KG}$ &
$\rho^{-2}\chi_0$ & $\rho^{-3}\Phi^{EM}_0$ &
$\rho^{-4}\Sigma^{RS}_0$ & $\rho^{-5}\Psi_0$ \\
$s$    &
-2 & -\nicefrac{3}{2} &
-1 & \nicefrac{1}{2} &
0 &
\nicefrac{1}{2} & 1 &
\nicefrac{3}{2} & 2 \\
GHP weight &
(-4,0) & (-3,0) & (-2,0) & (-1,0) & (0,0)& (1,0)& (2,0) & (3,0) & (4,0) \\
\end{tabular}
\end{ruledtabular}
\label{tab:F}
\end{table*}

\begin{widetext}
\begin{multline}
\frac{1}{G(y)}\left[ \frac{\hat{\Phi}_{,\tau\tau}}{\Ktau^2}+\frac{s\,G(y)_{,y}\hat{\Phi}_{,\tau}}{\Ktau}-\frac{\left(s+1\right)\left(2s+1\right)}{6}\,G(y)G(y)_{,yy}\hat{\Phi}\right] - \frac{\left[ G(y)^{s+1}\hat{\Phi}_{,y} \right]_{,y}}{G(y)^s} \\
+\frac{1}{G(x)}\left[ \frac{\hat{\Phi}_{,\phi\phi}}{\Kphi^2}-\frac{isG(x)_{,x}\hat{\Phi}_{,\phi}}{\Kphi}-\frac{s^2G(x)_{,x}^2\hat{\Phi}}{4} \right] + \Big[ G(x)\hat{\Phi}_{,x} \Big]_{,x} +\frac{s^2+\frac{1}{2}}{3}\, G(x)_{,xx}\hat{\Phi} = 0\,.
\label{eq:meq}
\end{multline}
\end{widetext}

This equation, which is a generalization of the Teukolsky master equation for the C\,--\,metric (yet, contrary to the Teukolsky equation itself, only for non-rotating case), can be separated by the following ansatz
\begin{equation}
\hat{\Phi} = e^{-i\omega t} e^{im\phi} \mathcal{Y}(y)\mathcal{X}(x)\,, 
\label{eq:sep-ansatz}
\end{equation}
which leads to equation for angular part \re{gSWSH} in the next section and for equation for radial part \re{sepY} in Section \ref{sec:radial}.


Only massless Klein\,--\,Gordon equation is separable on the C\,--\,metric background; and so is the Dirac equation --- which in the massless limit is the Weyl neutrino equation and thus is already included above.

We have also discussed only NP scalars of maximal spin weight because only for these the equations are decoupled in type D spacetimes. Of course, to solve completely, for example, the electromagnetic field, we would have to analyze also $\Phi_1$.

\section{Accelerated spin weighted spherical harmonics}\label{sec:ang}
According to \cite{PRD.73} the spin weighted spheroidal harmonics are regular solutions to the equation
\begin{multline}
\left[ \left( 1-x^2 \right) S^{(s)}_{(lm),x} \right]_{,x} + \Biggl[ \left( cx -s \right)^2 -s\left( s-1 \right)\\ + A^{(s)}_{(lm)} - \frac{\left( m+sx \right)^2}{1-x^2} \Biggr] S^{(s)}_{(lm)} = 0
\label{eq:SWSH}
\end{multline}
The constant $c$ arises in the equation during the separation of variables of various fields on the Kerr\,--\,Newman background only if $a\omega \neq 0$. In the limit $c=0$ the separation constant is $A^{(s)}_{(lm)}=l(l+1)-s(s+1)$ and then the standard spin weighted spherical harmonics are obtained.

The separation of variables of master equation for the C\,--\,metric \re{meq} leads to the equation 
\begin{multline}
\left[ G(x) \mathcal{X}^{(s)}_{(lm),x} \right]_{,x} +\Biggl[  \frac{s^2\!+\frac{1}{2}}{3} \; G(x)_{,xx} + \Lambda^{(s)}_{(lm)} \\- \frac{\left( \hat{m} -\frac{s}{2}\,G(x)_{,x}\right)^2}{G(x)}\Biggr] {\mathcal{X}}^{(s)}_{(lm)} = 0\,,\quad \hat{m}\equiv \frac{m}{K_\phi}\,, \label{eq:gSWSH}
\end{multline}
for angular part which, according to Sturm\,--\,Liouville theory, posses an infinite number of regular orthogonal solutions. These solutions form a base for $L^2$ functions on interval $x\in\langle -1,\,1\rangle$. This equation is a generalization of \re{SWSH} with $c=0$ (we are investigating non-rotating case, i.e. $a=0$) for accelerated sources\footnote{Notice that this is a general relativistic effect, in the limit $\Rp=\Rm=0$ and $\Kphi=1$ the standard spin weighted spherical harmonics are recovered.}.

If the function $G(\xi)$ is even ($G(\xi)=G(-\xi)$) and if $\mathcal{X}^{(s)}_{(lm)}(x)$ is solution of equation \re{gSWSH} then the solution for $-s$ is given by $\mathcal{X}^{(-s)}_{(lm)} (x) = \mathcal{X}^{(s)}_{(lm)}(-x)$.

With the help of computer algebra systems we are able to find a solution of \re{gSWSH} for arbitrary $s,\,m,\,\Lambda$ and general structure function $G(\xi)=\left( 1-\xi^2 \right)\left( 1+A\Rp\xi \right)\left( 1+A\Rm\xi \right)$ in terms of general Heun function. For some special cases we can find the solution explicitly, even in terms of rational functions, not only formally as a Heun general function.

\subsection{Extremal case}
For the extremal case $\Rp=\Rm$ the solution of \re{gSWSH} regular in the entire interval $x\in\langle -1,\,1\rangle$ with $m=0$ for general half-integer $s$ has been found in terms of rational functions (in this case, the hypergeometric function reduces to polynomial)
\begin{multline}
\mathcal{X}^{(s)}_{(l0)} = C^{(s)}_{(l0)} \frac{\left( 1-x^2 \right)^{\frac{1}{2}\,s}}{\left( 1+A\Rp x \right)^{1+s}}\times \\  \pFq{s-l}{s+l+1}{1+s}{\frac{ 1+A\Rp}{2}\frac{1+x}{1+A\Rp x}}\,,
\end{multline}
with the eigenvalues
\begin{equation}
\Lambda^{(s)}_{(l0)} = \left( 1-A^2\Rp^2 \right)\left[l\left( l+1 \right) + \frac{1}{3}\left( 1-s^2 \right)\right].
\end{equation}

Solutions of \re{gSWSH} with $\Rm=\Rp$ and $m\neq 0$ is given in terms of Heun confluent function with an unknown eigenvalues. But in the Minkowski limit $\Rp\rightarrow 0$ and $\Kphi=1$, these solutions go over to (valid for $m\geq 0$) 
\begin{multline}
S^{(s)}_{(lm)} =  \left( \frac{1-x}{1+x} \right)^{s/2} \left( 1-x^2 \right)^{m/2} \times \\ \pFq{-l+m}{l+m+1}{m-s+1}{\frac{1}{2}\left( 1+x \right)}\,,
\label{eq:hgY}
\end{multline}
where $l,\,s,\,m$ are from either $\mathbb{N}_0$ or $\mathbb{N}_0 + \nicefrac{1}{2}$. Up to normalization, these solutions are standard spin weighted spherical harmonics, which are given in terms of the Wigner $d$-functions (see \cite{jorg}, \cite{lgW})
\begin{equation}
Y^{(s)}_{(lm)}(x,\,\phi) = \sqrt{\frac{2l+1}{4\pi}}\, e^{im\phi} d^l_{-s,m} (x)\,,
\end{equation}
where 
\begin{multline}
d^l_{sm}(x) = \sum_{k=\max (0,\,s-m)}^{\min (l+s,\,l-s)}
\left( -1 \right)^{k-s+m}\, \times \\
\frac{ \sqrt{(l+s)!(l-s)!(l+m)!(l-m)!} }{ k!(l+s-k)!(l-k-m)!(k-s+m)! }\,\times\\
\left( \frac{\sqrt{1+x}}{2} \right)^{2l-2k+s-m}
\left( \frac{\sqrt{1-x}}{2} \right)^{2k-s+m}\,,
\label{eq:WigD}
\end{multline}
in which we trivially replaced $\cos\frac{\theta}{2}=\sqrt{\frac{1+x}{2}}$ and $\sin\frac{\theta}{2}=\sqrt{\frac{1-x}{2}}$.
 This gives us a relation (up to the normalization) of Wigner $d$-function and hypergeometric function \re{hgY}.

\subsection{General case}
For a general case ($\Rp\neq\Rm$) the solutions of \re{gSWSH} (only for $m=0$) are are found in terms of the Heun functions $Hf_l$:
\begin{multline}
\mathcal{X}^{(s)}_{(l0)} =  \frac{1}{1+A\Rm x}\left(\frac{1-x^2}{\left( 1+A\Rp x \right)\left( 1+A\Rm x \right)}\right)^{\frac{1}{2}\,s} \times \\
 H\!f_l\left(\vec{a},\,\vec{q};\,1,\,1-s,\,1-s,\,1+s;\,\frac{1+A\Rm}{2}\frac{1+x}{1+A\Rm x} \right) 
\end{multline}
where
\begin{align}
\vec{a} &= \frac{1}{2}\frac{\left(1-A\Rp  \right)\left( 1+A\Rm \right)}{A\left( \Rp-\Rm \right)},\\ 
\vec{q} &= \frac{1}{6}\frac{3\Lambda^{(s)}_{(l0)}-\left( s+1 \right) \left( 2s+1 \right)\left( 1-A^2\Rp\Rm \right)}{A\left( \Rp-\Rm \right)} + \frac{s+1}{2}\,,
\end{align}
with the eigenvalues given by
\begin{equation}
\Lambda^{(s)}_{(l0)} = \left[  l\left( l+1 \right)+\frac{1}{3}\left( 1-s^2 \right) \right]\sqrt{1-A^2\Rm^2}\sqrt{1-A^2\Rp^2}\,.
\end{equation}

\section{Scalar harmonics on ``sphere'' $(x,\,\phi)$}\label{sec:scalar}

The conformally rescaled 2-metric of constant $\tau$ and $y$ is 
\begin{equation}
\vec{\d} s^2_{(2D)} = \frac{\vec{\d} x^2}{G(x)}+\Kphi^2\,G(x)\,\vec{\d}\phi^2\,.
\end{equation}
The Laplace operator on this ``sphere'' reads
\begin{equation}
\Delta_{(2D)} \Phi(x,\phi) = \Bigl[ G(x) \Phi_{,x} \Bigr]_{,x} + \frac{1}{K_\phi^2}\frac{\Phi_{,\phi\phi}}{G(x)}\,,
\end{equation}
so the eigenvalue problem $\Delta_{(2D)}\Phi = -\Lambda_{(lm)}\Phi$ can be separated using ansatz $\Phi=e^{im\phi} \mathcal{\tilde{X}}$, where  
\begin{equation}
\left[ G(x) \mathcal{\tilde{X}}_{(lm),x} \right]_{,x} +\left[   \Lambda_{(lm)}- \frac{ \hat{m}^2}{G(x)}\right] {\mathcal{\tilde{X}}}_{(lm)} = 0\,. \label{eq:gX}
\end{equation}
Together with the appropriate boundary conditions this is again the formulation of Sturm\,--\,Liouville problem for eigenfunctions $\tilde{X}_{(lm)}$ and eigenvalues $\Lambda_{lm}$.

This equation (which arises from the separation of variables for the electromagnetic Ernst potential \re{PsiEq}) is clearly different from \re{gSWSH} (which arises during the separation of variables for master equation \re{meq}) because the latter contains a term proportional to $G(x)_{,xx}$ for any $s$. Thus, \re{gX} does not fit in the scheme immediately. Yet, it shows the way how to construct a basis of spin 1 weighted accelerated spherical harmonics from a scalar basis on this ``sphere'' $\tilde{Y}_{(lm)} = \tilde{\mathcal{X}}_{(lm)} e^{im\phi}$; the solutions of \re{gSWSH} are given by
\begin{align}
Y^{(-1)}_{(lm)} = \frac{\bar{\delta} \tilde{Y}_{(lm)}}{A(x-y)} &= \frac{ -\Kphi G(x)\,\tilde{Y}_{(lm),x}- i\tilde{Y}_{(lm),\phi}}{\Kphi\sqrt{G(x)}}\,, \\
& \left( \text{cf. } \bar{\mdh} = -\left( \frac{\p}{\p\theta}-\frac{i}{\sin\theta}\frac{\p}{\p\phi}  \right) \right); \nonumber \\
Y^{(1)}_{(lm)} = \frac{\delta \tilde{Y}_{(lm)}}{A(x-y)} &= \frac{ -\Kphi G(x)\,\tilde{Y}_{(lm),x}+ i\tilde{Y}_{(lm),\phi}}{\Kphi\sqrt{G(x)}}\,, \\
&\left( \text{cf. } \mdh = -\left( \frac{\p}{\p\theta}+\frac{i}{\sin\theta}\frac{\p}{\p\phi}  \right) \right); \nonumber
\end{align}
which resembles the standard procedure of generating spin\,--\,weighted spherical harmonics.

\section{Radial function}\label{sec:radial}
Separation of the master equation \re{meq}, furthermore, leads to the equation for radial function $\mathcal{Y}^{(s)}_{(lm)}$
\begin{multline}
\frac{\left[ G(y)^{s+1} \mathcal{Y}^{(s)}_{(lm),y} \right]_{,y}}{G(y)^{s}} + 
\Biggl[\frac{1}{G(y)}\Biggl( \omega^2-is\omega G(y)_{,y} \\ +\frac{\left( 2s+1 \right)\left( s+1 \right)}{6}\,G(y)\,G(y)_{,yy}\Biggr) + \Lambda^{(s)}_{(lm)} \Biggr]\mathcal{Y}^{(s)}_{(lm)} = 0.
\label{eq:sepY}
\end{multline}
Naturally there is a ``symmetry'' between solutions with $\pm s$. Let $\mathcal{Y}^{(|s|)}_{(lm)}$ be the solution of \re{sepY} with spin $|s|$. Then the function 
\begin{equation}
\mathcal{Y}^{(-|s|)}_{(ml)} = G(y)^{|s|} \mathcal{Y}^{(|s|)}_{(lm)}
\label{eq:Ys2}
\end{equation}
is the solution of \re{sepY} with spin $-|s|$ and frequency $-\omega$.

The equation \re{sepY} has the same structure of singular point as the structure function $G(\xi)$ and the infinity,
\begin{equation}
-\frac{1}{A\Rm},\,-\frac{1}{A\Rp},\,-1,\,1,\,\infty\,.
\end{equation}

The explicit solutions of \re{Ys2} is probably impossible to find but let us analyze the behavior of the solutions at the outer black hole horizon $y=-1/A\Rp$, the acceleration horizon $y=-1$ for the static case $\omega=0$. Then $y=-1/A\Rp$, $y=-1$ are regular singular points of the equation \re{Ys2} all of them with characteristic exponents $(0,\,-s)$ regardless the $\Lambda^{(s)}_{(lm)}$ for nondegenerated case $\Rp\neq\Rm$. In the extremal case\footnote{And thus only for axially symmetric configurations.} the characteristic exponents of singular point $y=-1/A\Rp$ are $(l-s,\,-l-s-1)$. In the C\,--\,metric the infinity which is interesting is $\Scri^+$ which is not easy to describe in these coordinates and thus the asymptotic behaviour of fields at $\Scri$ will be discussed elsewhere.

The theory of ordinary differential equations can provide us with the behaviour of two linearly independent solutions in the vicinity of regular singular points. Sort the exponents at the singularity and thus define $R_1=\max \{0,-s\}$ and $R_2=\min \{0,-s\}$ so that $R_1\geq R_2$. Let $\sigma_j = 1+\sum_{k=1}^{\infty} a_{jk} (y-y_s)^k$; the first solution around a singular point $y_s$ is then 
\begin{equation}
\mathcal{Y}^{(s)A}_{(lm)} =  \left( y-y_s \right)^{R_1}\sigma_1,  \\
\end{equation}
and the second one is for $R_1-R_2 \in \mathbb{N}_0+\nicefrac{1}{2}$
\begin{equation}
\mathcal{Y}^{(s)B}_{(lm)} = \left( y-y_s \right)^{R_2}\sigma_2 \,,  \\
\end{equation}
or, for $s \in \mathbb{N}_0$
\begin{equation}
\mathcal{Y}^{(s)B}_{(lm)} = \left( y-y_s \right)^{R_2}\sigma_2  + c\ln\left( y-y_s \right) \mathcal{Y}^{(s)A}_{(lm)} \,, \\
\end{equation}
where the constant $c$ can be zero.

In the static ($\omega=0$), axisymmetric ($m=0$) --- i.e. for static (moving along $\vec{\xi_{\tau}}$) axisymmetric sources of test field --- and extremal ($\Rm=\Rp$) case the two linearly independent solutions can be found explicitly in terms of hypergeometric functions, for $s\geq 0$
\begin{align}
\mathcal{Y}^{(s)r}_{(l0)} &= \frac{\left( 1+y \right)^{l-s}}{\left( 1+A\Rp y \right)^{l+s+1}}  \times \nonumber \\
&\qquad \pFq{-l}{s-l}{-2l}{\frac{2}{1+A\Rp}\,\frac{1+A\Rp y}{1+y}} \,,\\
\mathcal{Y}^{(s)s}_{(l0)} &= \frac{\left( 1+A\Rp y \right)^{l-s}}{\left( 1+y \right)^{l+s+1}} \times \nonumber \\
&\pFq{l+1}{l+s+1}{2( l+1)}{\frac{2}{1+A\Rp}\,\frac{1+A\Rp y}{1+y}} \,.
\end{align}
The solution $\mathcal{Y}^{(s)r}_{(l0)}$ is rational function divergent at the black hole horizon (i.e. it is the physical outer solution) and the solution $\mathcal{Y}^{(s)s}_{(l0)}$ is divergent at the acceleration horizon (i.e. it is a physical inner solution; the divergence at the acceleration horizon means that there cannot exist any static sources above the acceleration horizon, which is a desirable property of the solution). 

\section{Electromagnetic field, a deeper analysis}\label{sec:elmag}
First of all let us remind that there is an almost forgotten reference in Bi\v{c}\'ak \cite{biel1} that for the D type spacetimes Fackerell and Ipser \cite{fip} gave a decoupled equation even for electromagnetic scalar $\Phi_1$ 
\begin{eqnarray}
\bigl[ \left( D+\epsilon+\bar{\epsilon}-\varrho-\bar{\varrho} \right)\left( \Delta+2\mu \right)\qquad\qquad\qquad \label{eq:Phi1} \\- \left( \delta +\bar{\pi} -\bar{\alpha}+\beta-\tau\right)\left( \bar{\delta}+2\pi \right) \bigr]\Phi_1 &=&0\,, \nonumber
\end{eqnarray}
This equation is separable, using the ansatz
\begin{equation}
\Phi_1 = A^2(x-y)^2 e^{-i\omega \tau} e^{im\phi} \mathcal{Y}^{(0)}_{(lm)}(y) \mathcal{X}^{(0)}_{(lm)}(x)\,,
\label{eq:Phi1sep}
\end{equation}
the separation leads to equations
\begin{eqnarray}
\left( G(x)\mathcal{X}^{(0)}_{(lm),x} \right)_{,x} + \left[  \Lambda^{(0)}_{(lm)} - \frac{m^2}{\Kphi^2 G(x)} \right]\mathcal{X}^{(0)}_{(lm)} & = & 0\,, \label{eq:Phi1X} \\
\left( G(y)\mathcal{Y}^{(0)}_{(lm),y} \right)_{,y} + \left[ \Lambda^{(0)}_{(lm)} + \frac{\omega^2}{\Ktau^2 G(y)} \right]\mathcal{Y}^{(0)}_{(lm)} & = & 0\,, \label{eq:Phi1Y}
\end{eqnarray}
with eigenvalues
\begin{equation}
\Lambda^{(0)}_{(l0)} = l\left( l+1 \right)\sqrt{1-A^2\Rm^2}\sqrt{1-A^2\Rp^2}\,.
\end{equation}
In the equation \re{Phi1X} we can plainly recognize the equation \re{gX} for eigenfunctions of the Laplace operator on the ``sphere'' $(x,\,\phi)$.

The self-conjugated form of electromagnetic field tensor $\vec{F}^*$ can be reconstructed from the complex NP scalars in the case of non-rotating C\,--\,metric in the tetrad \re{PND}.

Le us take a closer look at the static electromagnetic field ($\omega=0$). 
The static observer, comoving with the black hole, has four velocity
\begin{equation}
\vec{u} = \frac{A(x-y)}{\sqrt{-G(y)}}\,\frac{\vec{\p}}{\vec{\p}\tau}\,.
\label{eq:obs}
\end{equation}
This observer measures electric field $\vec{E}$ and magnetic field $\vec{B}$
\begin{equation}
\vec{M} \equiv \vec{E}+i\vec{B} = \vec{u}\cdot\vec{F}^*\,,
\end{equation}
which we project onto the orthonormal basis $(\hat{\vec{e}}_y,\,\hat{\vec{e}}_x,\,\hat{\vec{e}}_\phi)$
\begin{align}
\hat{M}_y \equiv \vec{M}\cdot\hat{\vec{e}}_y &= -\Phi_{1} \,,\\
\hat{M}_x \equiv \vec{M}\cdot\hat{\vec{e}}_x & = \frac{1}{2} \frac{G(y)\Phi_0+A^2(x-y)^2\Phi_2}{A(x-y)\sqrt{-G(y)}} \,,\\
\hat{M}_\phi \equiv \vec{M}\cdot\hat{\vec{e}}_\phi &=\frac{i}{2} \frac{G(y)\Phi_0-A^2(x-y)^{2}\Phi_2}{A(x-y)\sqrt{-G(y)}} \,.
\end{align}
The scalar $\Phi_1$ describes purely radial behavior of the field and so it is not important for the discussion of the regularity conditions of the field in the vicinity of the axis.

Summing up all the known informations, i.e. ansatz from table \ref{tab:F} and ``symmetry'' in radial solutions \re{Ys2}, we get
\begin{align}
\Phi_0 &= a^{(1)}_{(lm)}{A^3(x-y)^3}{\mathcal{Y}^{(1)}_{(lm)}\mathcal{X}^{(1)}_{(lm)}}\,, \\
\Phi_2 &= a^{(-1)}_{(lm)}{A(x-y)}{\mathcal{Y}^{(-1)}_{(lm)}\mathcal{X}^{(-1)}_{(lm)}} \\
& = a^{(-1)}_{(lm)}{A(x-y)}{G(y)\mathcal{Y}^{(1)}_{(lm)}\mathcal{X}^{(-1)}_{(lm)}}\,,
\end{align}
thus, for the orthonormal tetrad components of electromagnetic field as measured by static observer we get
\begin{eqnarray}
M_x &=&  \frac{-1}{2}\, {A^2(x-y)^2}{\sqrt{-G(y)}\,\mathcal{Y}^{(1)}_{(lm)}}\,\times \nonumber \\ 
 &&\quad  \left( a^{(1)}_{(lm)}\mathcal{X}^{(1)}_{(lm)} + a^{(-1)}_{(lm)}\mathcal{X}^{(-1)}_{(lm)} \right) e^{im\phi} \,,\\
M_\phi &=&
  \frac{-1}{2}\, {A^2(x-y)^2}{\sqrt{-G(y)}\,\mathcal{Y}^{(1)}_{(lm)}}\, \times\nonumber \\    
 &&\quad  \left( a^{(1)}_{(lm)}\mathcal{X}^{(1)}_{(lm)} - a^{(-1)}_{(lm)}\mathcal{X}^{(-1)}_{(lm)} \right) e^{im\phi} \,,
\end{eqnarray}
for which the standard discussion (i.e. as done for Reissner\,--\,Nordstr\"{o}m black hole) of regularity of vector fields can be done.


In the section \ref{sec:proj} we showed how to find two solutions of test Maxwell equations using the Ernst formalism and provided some symmetries between equations for $m=0$. Let us use these result in the extreme and static case (so explicit solutions of radial and angular parts exist).

We can calculate the Ernst potentials $\Phi^{(\tau)}$ and $\Epsilon^{(\tau)}$ for the non-rotating C-metric. They are
\begin{eqnarray}
\Phi^{(\tau)} &=& y \,, \\
\Epsilon^{(\tau)} &=& \frac{G(y)}{\left( x-y \right)^2}-q^2y^2\,.
\end{eqnarray}
The appropriate NP scalars are given by \re{tau0}\,--\,\re{tau2}.

These solutions are static and axially symmetric therefore they can be expanded in terms of the basis given by solution of \re{Phi1X} and \re{Phi1Y} with $\omega=m=0$. They can be solved by looking for solution of $G\mathcal{U}_{,\xi\xi}+\Lambda\mathcal{U}$ as discussed in section \ref{sec:proj}. 

Let us investigate the extreme case $q=\Rm=\Rp$. Then we can find the solutions of \re{Phi1X} and \re{Phi1Y} in terms of hypergeometric functions
\begin{multline}
\mathcal{X}^{(0)r}_{(l0)} = \mathcal{Y}^{(0)r}_{(l0)} = \frac{\d}{\d \xi} \Biggl[ \left( 1-\xi \right)\left( \frac{1+A\Rp \xi}{1+\xi} \right)^{-l} \times \\
\pFq{-l}{1-l}{-2l}{\frac{2}{1+A\Rp}\frac{1+A\Rp \xi}{1+\xi}} \Biggr],
\end{multline}
which is regular solution in the interval $\xi\in\langle -1,\,1\rangle$ for $l\geq 1$ (for $l=0$ the solution is constant) and reduces to rational functions whereas
\begin{multline}
\mathcal{Y}^{(0)s}_{(l0)} = \frac{\d}{\d \xi} \Biggl[ \left( 1-\xi \right)\left( \frac{1+A\Rp \xi}{1+\xi} \right)^{l+1} \times \\
\pFq{2+l}{1+l}{2+2l}{\frac{2}{1+A\Rp}\frac{1+A\Rp \xi}{1+\xi}} \Biggr],
\end{multline}
is singular\footnote{Singular solutions for angular part are irrelevant.} at $\xi=-1$ and, due to the contiguous relations among hypergeometric functions (see \S 15.5(ii) of \cite{NIST}), can be expressed in terms of $\pFq{1}{1}{2}{z}$ which equals to $-z^{-1}\ln\left( 1-z \right)$ (a few formulas can be found in appendix A).

The trivial solution of Maxwell's equation, the ``Coulomb potential'', given by the Ernst potential $\Psi^{(\tau)}=\Phi^{(\tau)}=y$ leads to NP coefficients $\Phi_0=0$, $\Phi_1=-A^2\left( x-y \right)^2\sqrt{2}/\Ktau$ and $\Phi_2=0$ which compared with \re{Phi1sep} trivially yield expansion as $\Phi_1 \sim \mathcal{X}^{(0)r}_{(00)}\mathcal{Y}^{(0)r}_{(00)r}$.

The other solution of test Maxwell equations which is given by $\Psi^{(\tau)}=\Epsilon^{(\tau)}$ represents, in the Minkowski limit, uniform electric field aligned parallel to the symmetry axis. From \re{tau1} and \re{Phi1sep} follows
\begin{equation}
\Psi^{(\tau)}_{,y} = \frac{\d}{\d y}\left[ \frac{G(y)}{\left( x-y \right)^2} - \Rp^2 y^2\right] = \sum_{l=0}^{\infty} a^{(0)}_{(l0)} \mathcal{X}^{(0)r}_{(l0)} \mathcal{Y}^{(0)s}_{(l0)}\,.
\end{equation}
Firstly, we notice, that even the ``Coulomb part'' is present in the expansion. Secondly, although it looks nontrivial, it can be explicitly checked that the integral $\int_{-1}^{1} \left[ \frac{G(y)}{\left( x-y \right)^2} - \Rp^2 y^2\right] \mathcal{X}^{(0)r}_{(l0)}\,\d x$ (including logarithmic terms) is proportional to $\mathcal{Y}^{(0)s}_{(l0)}$ after using the contiguous relations for hypergeometric function. 

For this test field the NP scalars diverge at $y=-1$ because the ``sources'' are located at infinity. Yet, it is merely an artefact of the choice of tetrad, the invariant $\Phi_0\Phi_2-\Phi_1^2$ doesn't have any singularities at the acceleration horizon.

Clearly, the integral $\int_{-1}^1 \Phi_1/\left( x-y \right)^2\,\d x$ can be performed only for $y<-1$ because the range of the coordinates is given by the condition $x-y<0$ and thus the ``sphere'' $y=$const exists only for $y<-1$.

The series is infinite even in the Minkowski limit which only seemingly contradicts the known result that the homogeneous field is just an $l=1$ mode. The reason lies in the fact that this system is uniformly accelerated and therefore also even $\Phi_0$ and $\Phi_2$ are nonzero.

The detail analysis of electromagnetic field will be given in a following paper.

\section{Spin weight raising and lowering operators}\label{sec:sw}
From what we have seen so far it is necessary, contrary to the standard spin weighted harmonics, to distinguish between fields of different spin. For every spin $S$, there is a sequence\footnote{Therefore it would be more appropriate to denote these spin weighted spherical harmonics with ${Y}^{(s,S)}_{(lm)}$, or they $x$ component with $\mathcal{X}^{(s,S)}_{(lm)}$ --- we have not used this notation in the previous text. There it would be just a complication; we omitted the spin $S$.} of spin weighted accelerated spherical harmonics $Y^{(s,S)}_{(lm)}=\mathcal{X}^{(s,S)}_{(lm)}e^{im\phi}$ with spin weight $s\in(-S,\,-S+1,\dots,\,S)$. This sequence can be generated using the universal spin weight raising operator acting on spin weight component $s$
\begin{multline}
{Y}^{(s+1,S)}_{(lm)} = \mdh {Y}^{(s,S)}_{(lm)} = \\= \left( G(x) \right)^{\frac{s}{2}} \left( \frac{\Kphi G(x) \frac{\p}{\p x} -i\frac{\p}{\p\phi} }{\sqrt{G(x)}}\right) \left( Y^{(s,S)}_{(lm)} \left( G(x) \right)^{-\frac{s}{2}} \right),
\label{eq:Raising}
\end{multline}
and spin weight lowering operator
\begin{multline}
{Y}^{(s-1,S)}_{(lm)} = \bar{\mdh} {Y}^{(s,S)}_{(lm)} = \\= \left( G(x) \right)^{-\frac{s}{2}} \left( \frac{\Kphi G(x) \frac{\p}{\p x} +i\frac{\p}{\p\phi} }{\sqrt{G(x)}}\right) \left( Y^{(s,S)}_{(lm)} \left( G(x) \right)^{\frac{s}{2}} \right).
\label{eq:Lowering}
\end{multline}
The complete scheme can be found in table \ref{tab:S}, where the arrows mean lowering and raising spin weight and the fields are labeled by their spin and spin weight (for example gravitational perturbation $\Psi_4 \rightarrow \Psi^{(-2,2)}$ or $\Sigma_3 \rightarrow \Sigma^{(-\nicefrac{3}{2},\nicefrac{3}{2})}$ and so on).
In the limit $\Rp\rightarrow 0$ and $\Kphi\rightarrow 1$, the eigenfunctions of operators no longer depend on spin of the field $S$ but only on the spin weight $s$ of the particular component (as schematically depicted in the last two lines ot the table \ref{tab:S}).

Thus, more properly, the equation \re{gSWSH} for accelerated spin weighted spherical harmonics should be written as
\begin{multline}
\left[ G(x) \mathcal{X}^{(\pm S,S)}_{(lm),x} \right]_{,x} +\Biggl[  \frac{S^2\!+\frac{1}{2}}{3} \; G(x)_{,xx} + \Lambda^{(\pm S,S)}_{(lm)} \\- \frac{\left( \hat{m} -\frac{(\pm S)}{2}\,G(x)_{,x}\right)^2}{G(x)}\Biggr] {\mathcal{X}}^{(\pm S,S)}_{(lm)} = 0\,,\quad \hat{m}\equiv \frac{m}{K_\phi}\,; \label{eq:GSWSH2}
\end{multline}
it holds only for the maximal spin weight $s=\pm S$.

Although the equations for non\,--\,extreme components ($|s|\neq S$) are not known for $S>1$, the following statement can be proved: Let ${Y}^{(S,S)}_{(lm)}$ be a solution of \re{GSWSH2} for arbitrary $S$. Then applying $(2S-1)$ times $\bar{\mdh}$ and thus obtaining ${Y}^{(-S,S)}_{(lm)}$ we get a solution of eigenvalue problem \re{GSWSH2} with $s=-S$ if
\begin{equation}
\frac{\d^n}{\d\xi^n}\,G(\xi) = 0\qquad\text{for} \qquad n\geq 5\,,
\end{equation}
which holds even for the charged and rotating C\,--\, metric.

\newcommand{\arw}{$\longleftrightarrow$}
\begin{table*}
\centering
\caption{Accelerated spin weighted spherical harmonics schema}
\renewcommand{\arraystretch}{1.5}
\begin{ruledtabular}
\begin{tabular}{l*{9}{c}} 
\tikzmark{a} 
& 
$\Psi^{(-2,2)}$ & 
\arw & 
$\Psi^{(-1,2)}$ & 
\arw & 
$\Psi^{(0,2)}$ &
\arw & 
$\Psi^{(1,2)}$ &
\arw &
$\Psi^{(2,2)}$ \\
\parbox[t]{8mm}{\multirow{4}{*}{\rotatebox[origin=c]{90}{$\Rp\rightarrow 0,\,\Kphi \rightarrow 1$}}}&
&
$\Sigma^{(-\nicefrac{3}{2},\nicefrac{3}{2})}$ &
\arw &
$\Sigma^{(-\nicefrac{1}{2},\nicefrac{3}{2})}$ &
\arw &
$\Sigma^{(\nicefrac{1}{2},\nicefrac{3}{2})}$ &
\arw &
$\Sigma^{(\nicefrac{3}{2},\nicefrac{3}{2})}$ & 
\\
 &
&
&
$\Phi^{(-1,1)}$ &
\arw &
$\Phi^{(0,1)}$ &
\arw &
$\Phi^{(1,1)}$ &
&
\\
 &
&
&
&
$\chi^{(-\nicefrac{1}{2},\nicefrac{1}{2})}$ &
\arw &
$\chi^{(\nicefrac{1}{2},\nicefrac{1}{2})}$ &
&
&
\\
 & 
&
&
&
&
$\Phi^{(0,0)}$ &
&
&
&
\\ \cline{2-10}
\tikzmark{b}
& 
$\Phi^{(-2)}$ & 
\arw & 
$\Phi^{(-1)}$ & 
\arw & 
$\Phi^{(0)}$ &
\arw & 
$\Phi^{(1)}$ &
\arw &
$\Phi^{(2)}$ \\
&
&
$\Sigma^{(-\nicefrac{3}{2})}$ &
\arw &
$\Sigma^{(-\nicefrac{1}{2})}$ &
\arw &
$\Sigma^{(\nicefrac{1}{2})}$ &
\arw &
$\Sigma^{(\nicefrac{3}{2})}$ & 

\end{tabular}
\begin{tikzpicture}[overlay, remember picture, yshift=.25\baselineskip, shorten >=.5pt, shorten <=.5pt]
    \draw [->] ([yshift=.75pt]{pic cs:a}) -- ({pic cs:b});
  \end{tikzpicture}
\end{ruledtabular}
\label{tab:S}
\end{table*}

\section{Cosmic string spacetime}\label{sec:cs}

As soon as the spherical symmetry is abandoned everything becomes more complicated. As one of the simplest examples the cosmic string spacetime can serve.
The metric can be obtained as a zero mass and zero charge limit of the C\,--\,metric; written in quasi\,--\,spherical coordinates it reads
\begin{equation}
\vec{\d} s^2 = -\vec{\d} t^2 + \vec{\d} r^2 +r^2\left( \vec{\d}\theta^2 + \Kphi^2\sin^2\theta\,\vec{\d}\phi^2\right) \,,
\end{equation}
and represents flat spacetime with a cosmic string aligned along the $z$-axis.

To analyse fields on this spacetime it is evidently of no sense to use spherical harmonics.

The relevant (angular) part of the metric is
\begin{multline}
\vec{\d} s^2 = \vec{\d}\theta^2 + \Kphi^2\sin^2\theta\,\vec{\d}\phi^2 \\ 
= \frac{\vec{\d} x^2}{1-x^2}+(1-x^2)\,\Kphi^2\,\vec{\d}\phi^2 \,.
\end{multline}
The separation of variables for equation $\Delta \Phi(x,\phi) = -\lambda\Phi(x,\phi)$ leads to the ansatz $\Phi(x,\phi)=Y^{(s)}_{(lm)}(x,\phi)= S^{(s)}_{(lm)}(x)\, e^{im\phi}$ where
\begin{equation}
\frac{1}{S^{(0)}_{(lm)}}\frac{\d}{\d x} \left[ \left( 1-x^2 \right) \frac{\d S^{(0)}_{(lm)}}{\d x} \right] - \frac{\left(\frac{m}{\Kphi}\right)^2}{1-x^2} -\frac{1}{3} + \Lambda^{(s)}_{(lm)} = 0\,,
\end{equation}
which can be considered as limit case of \re{gSWSH} (with $\Rp=\Rm=0$, $s=0$) and therefore we will solve the more general case with an arbitrary half-integer $s$:
\begin{multline}
\frac{1}{S^{(s)}_{(lm)}}\frac{\d}{\d x} \left[ \left( 1-x^2 \right) \frac{\d S^{(s)}_{(lm)}}{\d x} \right] - \frac{\left(\frac{m}{\Kphi}+sx  \right)^2}{1-x^2} 
\\ - \frac{2}{3}\,s^2-\frac{1}{3} + \Lambda^{(s)}_{(lm)} = 0\,.
\end{multline}
The general regular solution reads
\begin{multline}
S^{(s)}_{(lm)} = \left( 1-x \right)^{l+m\frac{1-\Kphi}{\Kphi}} \left( \frac{1+x}{1-x} \right)^{\frac{1}{2}\left(\frac{1}{\Kphi}\,m-s\right)} \times \\ 
\pFq{-l+m}{-l-s-m\,\frac{1-\Kphi}{\Kphi}}{1+\frac{m}{\Kphi}-s}{-\frac{1+x}{1-x}}
\end{multline}
or, equivalently,
\begin{multline}
S^{(s)}_{(lm)} = \left( 1-x \right)^{\frac{m}{\Kphi}}
\left( \frac{1+x}{1-x} \right)^{\frac{1}{2}\left( \frac{m}{\Kphi}-s \right)}\times\\
\mathrm{JacobiP}\left( l-m, \frac{m}{\Kphi}+s,\frac{m}{\Kphi}-s,x \right)
\end{multline}
with eigenvalues
\begin{equation}
\Lambda^{(s)}_{(lm)} = \left( l+m\,\frac{1-\Kphi}{\Kphi} \right)\left( l+1+m\,\frac{1-\Kphi}{\Kphi} \right) + \frac{1}{3}\left( 1-s^2 \right).
\label{eq:csegv}
\end{equation}
The solutions are bounded if $s,\,l,\,m$ and $\Kphi$ fulfill following conditions: $\Lambda^{(s)}_{(lm)}-\frac{2}{3}\,s^2-\frac{1}{3}\,s>0$ and $|\frac{m}{K_\phi}|\geq |s|$, in other cases $S^{(s)}_{(lm)}=0$.

In the limit $\Kphi \rightarrow 1$ the standard spin weighted spherical harmonics\footnote{Euler's transformations of hypergeometric function, as well as Heun general functions, are used to change the form to a equivalent one (cf. \cite{NIST}).} are recovered.

We can see that $m$ enters the eigenvalues \re{csegv} even in this trivial case, therefore we have to expect that they enter eigenvalues of the accelerated spherical harmonics in a more complicated way.

\section{Conclusions}
We have derived an analogy of the Teukolsky master equation for non-rotating C\,--\,metric; provided the notion of accelerated spin weighted spherical harmonics, some of which we have found explicitly. We paid special attention to electromagnetic field which we solved completely. We showed that for non-axisymmetric configurations $m\neq 0$ the complete basis is difficult to find.

\begin{acknowledgements}
D.K. acknowledges the support from the Czech Science Foundation, Grant No. 14-37086G --- the Albert Einstein Centre. Moreover, D.K. would like to thank Prof. J. Bi\v{c}\'ak for introducing him to the C\,--\,metric, to Dr. M. Scholtz for carefully reading the manuscript and to Dr. P. Krtou\v{s} and Prof. J. Podolsk\'{y} for inspiring questions and comments.
\end{acknowledgements}

\appendix
\section{The Gauss contiguous relations}
The Gauss contiguous relations \cite{NIST} allows us to reduce some of the hypergeometric functions to elementary functions using the fact that 
\begin{equation}
L \equiv \pFq{1}{1}{2}{z} = -\frac{\ln\left( 1-z \right)}{z}\,.
\end{equation}
We for a general hypergeometric function $W \pFq{a}{b}{c}{z}$ we can define the operators $A^+$, $B^+$ and $C^+$ which are raising the parameters $a$, $b$ and $c$ by
\begin{align}
\pFq{a+1}{b}{c}{z}& = A^+(W) = \frac{z}{a}\,\frac{\d W}{\d z} + W \,, \\
\pFq{a}{b+1}{c}{z}& = B^+(W) = \frac{z}{b}\,\frac{\d W}{\d z} + W \,, \\
\pFq{a}{b}{c+1}{z}& = C^+(W) \nonumber\\
 & \hspace{-2em}=\frac{\left( 1-z \right)c\,\frac{\d W}{\d z} - c\left( a+b-c \right)W}{\left( c-a \right)\left( c-b \right)}\,.
\end{align}
Using these operators the hypergeometric function for $a,\,b,\,c\in \mathbb{N}$ and $c\geq 2$ can be expressed in terms of logarithms by
\begin{equation}
\pFq{a}{b}{c}{z} = \left( A^+ \right)^{a-1}\left( B^+ \right)^{b-1}\left( C^+ \right)^{c-2} L\,.
\end{equation}

%


\end{document}